\documentclass[conference]{IEEEtran}
\IEEEoverridecommandlockouts
\usepackage{cite}
\usepackage{amsmath,amssymb,amsfonts}
\usepackage{algorithmic}
\usepackage{graphicx}
\usepackage{textcomp}
\usepackage{multirow}
\usepackage{xcolor}
\def\BibTeX{{\rm B\kern-.05em{\sc i\kern-.025em b}\kern-.08em
    T\kern-.1667em\lower.7ex\hbox{E}\kern-.125emX}}
\usepackage{acronym}
\usepackage{color}
\usepackage{url}
\usepackage[caption=false]{subfig}
\usepackage{flushend} 
\usepackage[citecolor=black,
            colorlinks=true,
            linkcolor=black,
            urlcolor  = black,
            pdftitle={},
            pdfauthor={George Close, Thomas Hain, 
            Stefan Goetze},
            pdfkeywords={},
            pdfsubject={}]{hyperref}
\usepackage[english]{babel}
\addto\extrasenglish{}
\addto\extrasenglish{}
\addto\extrasenglish{}
\addto\extrasenglish{}
\addto\extrasenglish{}
\addto\extrasenglish{}
\usepackage{orcidlink}

\acrodef{DPT}   {dual-path Transformer}
\acrodef{GAN}   {generative adversarial network}
\acrodef{UDASE} {unsupervised domain adaptation speech enhancement}
\acrodef{CMGAN} {conformer-based metric \ac{GAN}}
\acrodef{SSSR}  {Self Supervised Speech Representation}
\acrodef{BLSTM} {bi-di\-rec\-tional long short-term memory}
\acrodef{STFT}  {short-time Fourier transform}
\acrodef{ISTFT} {inverse short-time Fourier transform}
\acrodef{PESQ}  {Perceptual Evaluation of Speech Quality}
\acrodef{STOI}  {Short-Time Objective Intelligibility}
\acrodef{DNSMOS}{Deep Noise Suppression Mean Opinion Score}
\acrodef{MOS}   {Mean Opinion Score}
\acrodef{SI-SDR}{scale invariant signal-distortion ratio}
\acrodef{NISQA} {Non-Intrusive Speech Quality Assessment}
\acrodef{MOS}   {mean opinion score}
\acrodef{MUSHRA}{MUltiple Stimuli with Hidden Reference and Anchor}
\acrodef{CNN}   {Convolutional Neural Network}
\acrodef{ASR}   {automatic speech recognition}
\acrodef{SQ}    {speech quality}
\acrodef{RMSE}  {root mean squared error}
\acrodef{GRU}   {gated recurrent unit}
\acrodef{SE}    {speech enhancement}

\begin{document}

\title{Hallucination in Perceptual Metric-Driven\\ Speech Enhancement Networks\\
\thanks{This work was supported by the Centre for Doctoral Training in Speech and Language Technologies (SLT) and their Applications funded by UK Research and Innovation [grant number EP/S023062/1]. This work was also funded in part by TOSHIBA Cambridge Research Laboratory.}
}

\author{\IEEEauthorblockN{George Close$^{\orcidlink{0000-0002-9478-5421}}$, Thomas Hain$^{\orcidlink{0000-0003-0939-3464}}$, and Stefan Goetze$^{\orcidlink{0000-0003-1044-7343}}$}
\IEEEauthorblockA{\textit{Speech and Hearing Group},  
\textit{Department of Computer Science, University of Sheffield}, 
Sheffield, United Kingdom}
\texttt{\{glclose1, t.hain, s.goetze\}@sheffield.ac.uk}

}

\maketitle

\begin{abstract}
Within the area of speech enhancement, there is an ongoing interest in the creation of neural systems which explicitly aim to improve the perceptual quality of the processed audio. In concert with this is the topic of non-intrusive (i.e.~without clean reference) speech quality prediction, for which neural networks are trained to predict human-assigned quality labels directly from distorted audio. When combined, these areas allow for the creation of powerful new speech enhancement systems which can leverage large real-world datasets of distorted audio, by taking inference of a pre-trained speech quality predictor as the sole loss function of the speech enhancement system. This paper aims to identify a potential pitfall with this approach, namely hallucinations which are introduced by the enhancement system `tricking' the speech quality predictor.
\end{abstract}


\section{Introduction}
There has been an ongoing interest in the development of neural \emph{non-intrusive} \ac{SQ} predictors, for both, (i) replacing traditional signal-processing-based \emph{intrusive} \ac{SQ} metrics~\cite{Kumar2023TorchaudioSquimRS,fu2021metricganu} and (ii) prediction of human \ac{MOS} quality labels~\cite{NISQA,mittag20b_interspeech,Cauchi19Quality,yi2022conferencingspeech}. Such neural-network-based predictors have two major benefits: In the case of predictors of traditional metrics, they allow for the creation of non-intrusive versions of normally intrusive \ac{SQ} metrics, such as e.g.~the frequently-used \ac{PESQ}~\cite{pesq}. While traditional metrics are often non-differentiable, these neural-network-based metrics can then be used in a loss function for training an \ac{SE} system~\cite{fu2022metricgan,close2022}. Direct \ac{MOS} predictors estimating human signal quality assessment are also able to leverage large datasets of audio with \ac{MOS} labels as training data where the model is trained to accurately predict the \ac{MOS} of the input audio. During inference, these models are able to mimic the effect of actually conducting listening tests which is often a costly and time-consuming process. Inference of \ac{MOS} predictors can also be used as a training objective for \ac{SE} systems, potentially allowing for the leveraging of large amounts of \emph{real} noisy audio (where no reference signal exists) as training data. Approaches combining metric and \ac{MOS} prediction~\cite{close22_interspeech} have also proved successful, especially when a strong correlation between metric and human labels can be found. 

In the recent CHiME7 challenge \ac{UDASE} task~\cite{leglaive2023chime7,leglaive2024objective} it was shown that high metric scores from non-intrusive neural \ac{SQ} predictors do not always match with actual human \ac{MOS} evaluation. The evaluation of the \ac{SE} system entries to the \ac{UDASE} task had two stages; first the entries were evaluated in terms of the scores from the \ac{DNSMOS}~\cite{reddy2022dnsmos} neural non-intrusive \ac{SQ} metric. Then, in the second evaluation stage, listening tests were conducted and \ac{MOS} scores for audio enhanced by the challenge entries were computed from these listening tests. 
The best-performing system in the first evaluation stage was \cite{chime7entryclose}, an \ac{SE} system which utilises a non-intrusive MetricGAN~\cite{fu2021metricganu} framework to directly optimise towards the \ac{DNSMOS} metric. However, this system was scoring lowest of the entries going forward to the listening-test evaluation stage; by optimising directly for high \ac{DNSMOS} scores, the \ac{SE} system 
may learn to introduce specific distortions which result in 
high \ac{DNSMOS} scores but which negatively impact the actual perceptual quality of the enhanced audio when assessed by humans. 

In general, it was found in the \ac{UDASE} task results~\cite{leglaive2024objective} that quality ratings from non-intrusive quality predictors such as \ac{DNSMOS} and 
TorchAudio-SQUIM~\cite{Kumar2023TorchaudioSquimRS} did not correlate strongly with the \ac{MOS} ratings obtained in the second evaluation stage by listening tests and that traditional intrusive signal processing based metrics such as \ac{PESQ} and \ac{STOI} showed significantly stronger correlation. 

This work therefore has two major objectives. Firstly, to better understand how \ac{SE} systems like that in \cite{chime7entryclose} learn to optimise their outputs towards neural non-intrusive \ac{SQ} metrics during training. Secondly, to identify why neural non-intrusive \ac{SQ} metrics fail to properly assess the human assessed quality of the output of \ac{SE} systems, even in the setting that the \ac{SE} system does not directly optimise the metric in training. 

This paper is structured as follows. 
In \autoref{sect:predictor} a novel non-intrusive \ac{SQ} predictor system is introduced. Then in \autoref{sect:se_model} an  \ac{SE} system which uses inference of the \ac{SQ} predictor in its training loss function is detailed. In \autoref{sect:exp1}, an experiment training the \ac{SE} system with varying degrees of influence of the \ac{SQ} predictor is conducted, and the results analysed. In \autoref{sect:exp2} a small listening test experiment is carried out using the models trained in the proceeding section, and the results analysed. \autoref{sec:Conclusion} concludes the paper.

\section{Non-Intrusive Speech Quality Predictor}\label{sect:predictor}
The non-intrusive \ac{SQ} predictor $\mathcal{D}(\cdot)$ used in this work is based on~\cite{bas-xlsr} and consists of a Transformer~\cite{vaswani2023attention} block, followed by a feed-forward attention block with a sigmoid activation on a single output neuron which represents the predicted quality $\hat{q}'$ of the input audio, normalised between $0$ and $1$. 

The proposed predictor differs from that in~\cite{bas-xlsr} as follows: Rather than an input feature derived from the XLS-R representation, the input feature of $\mathcal{D}(\cdot)$ is the output of the Transformer Encoder stage of a pre-trained Whisper~\cite{whisper} \ac{ASR} network. This representation has been shown to be a useful feature representation for similar non-intrusive prediction tasks \cite{winnerCPC2paper,mogridge2024nonintrusive}.
In this work, the \texttt{whisper-small}\footnote{\texttt{\url{https://huggingface.co/openai/whisper-small}}} model, trained on $680$k hours of labelled speech data is used. The encoder stage of this model returns a representation of fixed dimension $F_{\mathrm{Enc}} \times T_{\mathrm{Enc}} = 768 \times 1500$. Note that the Whisper encoder block is used solely as a feature extractor, and its parameters are not updated during the training of $\mathcal{D}(\cdot)$.

The metric prediction network $\mathcal{D}(\cdot)$ is trained as follows: The \ac{MOS} label $q$ in most datasets is expressed as a value between $1$ and $5$, higher being better. In the training and inference of $\mathcal{D}(\cdot)$, this value is normalized between $0.2$ and $1$, which is denoted as $q'$. For a pair of audio and normalized MOS label $\{x[n], \, q'\}$, the model is trained with a loss between the output of the model (i.e.~the predicted quality of $x[n]$) and the true normalized MOS label $q'$:
\begin{equation}
    L_\mathcal{D} = (\mathcal{D}(x[n]) - q')^2
\end{equation}
The model is trained following a scheme similar to that proposed in \cite{NISQA} where training halts if the validation performance does not improve after $20$ epochs. 

The performance of the proposed non-intrusive metric prediction network $\mathcal{D}(\cdot)$, trained and tested on the NISQA~\cite{NISQA} dataset is shown in \autoref{tab:nisqa-results}, compared to the NISQA baseline. The NISQA test set and (baseline) model are widely used benchmarks for the \ac{SQ} prediction task. 
The proposed predictor network outperforms this baseline both in terms of  spearman correlation $r$ and \ac{RMSE} across all three NISQA testsets (P501, FOR and LIVETALK), and is comparable or better than state-of-the-art systems~\cite{bas-xlsr,msqat} on these testsets.
In addition, a variant of $\mathcal{D}(\cdot)$, denoted as $\mathcal{D_B}(\cdot)$ in \autoref{tab:nisqa-results}, is trained based on additional datasets, i.e ~NISQA~\cite{NISQA}, Tencent~\cite{yi2022conferencingspeech} and PTSN~\cite{mittag20b_interspeech} speech quality datasets, which shows similar, in mean further increased performance. 

\begin{table}[!ht]
\caption{Proposed \ac{SQ} Predictor compared with baseline NISQA model.}
\label{tab:nisqa-results}
\resizebox{\columnwidth}{!}{
\begin{tabular}{l|cc|cc|cc|cc}
\textbf{Testset}                          
& \multicolumn{2}{c|}{P501}     
& \multicolumn{2}{c|}{FOR}      
& \multicolumn{2}{c|}{LIVETALK} 
& \multicolumn{2}{c}{MEAN}       \\ \hline
\textbf{Model} 
& r$\uparrow$         & RMSE$\downarrow$       
& r$\uparrow$         & RMSE$\downarrow$       
& r$\uparrow$         & RMSE$\downarrow$       
& r$\uparrow$         & RMSE$\downarrow$       \\ 
NISQA & 0.89          & 0.46          & 0.88          & 0.40          & 0.70          & 0.67          & 0.82          & 0.51          \\
$\mathcal{D}(\cdot)$  & \textbf{0.94} & \textbf{0.35} & 0.93 & \textbf{0.32} & 0.81 & 0.54 & 0.89 & \textbf{0.40}\\
$\mathcal{D}_B(\cdot)$ & 0.93	&0.37	&\textbf{0.94}	&0.32	&\textbf{0.85}	& \textbf{0.50}	& \textbf{0.91}	& \textbf{0.40}
\end{tabular}
}
\end{table}

\section{Speech Enhancement System}\label{sect:se_model}
The DPT-FSNet~\cite{dpt-fsnet} single-channel speech enhancement architecture which is based on the \ac{DPT} architecture is used as the baseline speech enhancement system denoted as $\mathcal{G}(\cdot)$ in this work. This model has shown state-of-the-art performance in this task, despite a relatively small parameter count.  It takes as input the real and imaginary \ac{STFT} components $\mathbf{X}_r$ and $\mathbf{X}_i$ of the noisy time domain signal $x[n]$, and returns mask matrices $\mathbf{M}_r$ and $\mathbf{M}_i$ which are multiplied with the inputs to produce estimated of the clean complex signal spectra, i.e.~$\mathbf{\hat{S}}_r$ and $\mathbf{\hat{S}}_i$. These are then used as inputs to an \ac{ISTFT} operation to produce the enhanced time domain audio $\hat{s}[n]$. For a detailed description of the architecture see \cite{dpt-fsnet}.

\subsection{Network Structure}
The DPT-FSNet has three sequential stages. The first stage is an encoder comprised of a $1$-D convolutional layer followed by a dilated-dense block with $4$ dilated convolutional layers. The input to the encoder are the real and imaginary \ac{STFT} components of the noisy input signal $x[n]$ in the form $2 \times T \times F$, and the output is a higher dimensional representation with dimensions $64 \times T \times F$, i.e.~$64$ time-frequency maps. 

The second stage comprises $4$ dual path \cite{9054266} Transformer blocks, which operate over the $T$ and $F$ dimensions sequentially. The Transformer blocks use a slight variation on the standard structure; i.e.~a \ac{GRU} layer is incorporated into the feed-forward network following the multi-head attention. 

The final stage is a decoder structured as a mirror of the initial encoder stage; it takes as input the output of the second stage and returns predicted real and imaginary masks $\mathbf{M}_r,\mathbf{M}_i$ with dimensions $2 \times T \times F$.

\subsection{Loss Function}
To train the proposed adaptation of the DPT-FSNet network $\mathcal{G}(\cdot)$, an extended loss function 
\begin{equation}
L = \alpha L_\mathrm{Spec} + (1-\alpha)L_\mathrm{SQ} 
\label{eq:joint_loss}
\end{equation}
is proposed which adds a loss term 
\begin{equation}
    L_\mathrm{SQ} = (1 - \mathcal{D}_B(\hat{s}[n]))^2
\label{eq:sq_loss}
\end{equation} 
based on inference of the  non-intrusive pre-trained \ac{SQ} predictor (cf.~\autoref{sect:predictor}) of the enhanced audio $\hat{s}[n]$ to the loss used in the original DPT-FSNet paper~\cite{dpt-fsnet}
\begin{equation}
\begin{split}
        L_\mathrm{spec} = \frac{1}{TF}\sum_{t=0}^{T-1}\sum_{f=0}^{F-1}\left|\left(|\mathbf{S}_{r}[t,f]| 
    - |\mathbf{\hat{S}}_{r}[t,f]|\right)  + \right. \\
    \left. \left(|\mathbf{S}_{i}[t,f]| - |\mathbf{\hat{S}}_{i}[t,f]|\right)\right|,
\end{split}
\label{eq:spec_loss}
\end{equation}
which is based on the distance of enhanced real and imaginary spectrogram components $\mathbf{\hat{S}}_r$ and $\mathbf{\hat{S}}_i$ to the spectrogram components $\mathbf{S}_{r}$ and $\mathbf{S}_{i}$ of the reference audio $s[n]$. Note that the time domain loss term as outlined in~\cite{dpt-fsnet} is not utilised here.
\begin{table*}[!t]
\caption{Performance of Speech Enhancement for different $\alpha$ in (\ref{eq:joint_loss}) for the VoiceBank-DEMAND testset.} Best performance denoted in \textbf{bold} font. Unprocessed data denoted in \textit{italic} font.
\label{tab:results1}
\centering
\begin{tabular}{c|c|cccccc|cccc}
\centering
\textbf{} & \textbf{}     & \textbf{} & \textbf{} & 
\multicolumn{3}{c}{\textbf{Composite}}  & \textbf{} & \multicolumn{3}{c}{\textbf{DNSMOS}}& \textbf{} \\
\multirow{-2}{*}{\textbf{Loss}} & \multirow{-2}{*}{\textbf{$\alpha$} \textbf{in (\ref{eq:joint_loss})}}     & \multirow{-2}{*}{\textbf{PESQ}} & \multirow{-2}{*}{\textbf{STOI}} & \textbf{CSIG} & \textbf{CBAK} & \textbf{COVL} & \multirow{-2}{*}{\textbf{SISDR}} & \textbf{SIG} & \textbf{BAK} & \textbf{OVR} & \multirow{-2}{*}{\textbf{$\mathcal{D}_B(\cdot)$}} \\ \hline
\multicolumn{1}{c|}{-} & \textit{clean} &\textit{4.50}&\textit{1.00}&	\textit{5.00}&	\textit{5.00}&	\textit{5.00}&	\textit{91.14}&	\textit{4.27}&	\textit{4.36}&\textit{3.88}&	\textit{0.67}\\
\multicolumn{1}{c|}{-} & \textit{noisy} & \textit{1.97}          & \textit{0.92}          & \textit{3.34}          & \textit{2.44}          & \textit{2.63}          & \textit{8.44}           & \textit{\textbf{4.24}}       & \textit{3.32}                & \textit{3.36}                & \textit{0.58}             \\\hline
\multicolumn{1}{c|}{$L_\mathrm{spec}$ only, (\ref{eq:spec_loss})}  & 1              & \textbf{2.99} & \textbf{0.95} & \textbf{4.09} & \textbf{3.57} & \textbf{3.55} & \textbf{19.82} & 4.14                & 4.42                & 3.86                & 0.65             \\
\multicolumn{1}{c|}{} & 0.9            & 2.93          & 0.94          & 3.96          & 3.52          & 3.44          & 19.70          & 4.12                & \textbf{4.46}       & \textbf{3.95}       & 0.68             \\
\multicolumn{1}{c|}{} & 0.8            & 2.63          & 0.93          & 3.59          & 3.33          & 3.09          & 19.62          & 4.05                & 4.41                & 3.91                & 0.70             \\
\multicolumn{1}{c|}{} & 0.7            & 2.72          & 0.94          & 3.78          & 3.38          & 3.24          & 19.39          & 4.08                & 4.30                & 3.78                & 0.71             \\
\multicolumn{1}{c|}{} & 0.6            & 2.63          & 0.93          & 3.45          & 3.25          & 3.00          & 19.25          & 4.00                & 4.33                & 3.79                & 0.79             \\
\multicolumn{1}{c|}{} & 0.5            & 2.65          & 0.93          & 3.57          & 3.29          & 3.07          & 19.43          & 4.04                & 4.36                & 3.84                & 0.77             \\
\multicolumn{1}{c|}{} & 0.4            & 2.66          & 0.93          & 3.67          & 3.31          & 3.14          & 18.92          & 4.06                & 4.28                & 3.75                & 0.76             \\
\multicolumn{1}{c|}{} & 0.3            & 2.68          & 0.93          & 3.79          & 3.34          & 3.22          & 18.98          & 4.11                & 4.38                & 3.83                & 0.77             \\
\multicolumn{1}{c|}{} & 0.2            & 2.58          & 0.93          & 3.47          & 3.25          & 3.00          & 18.51          & 3.92                & 4.24                & 3.70                & 0.75             \\
\multicolumn{1}{c|}{\multirow{-9}{*}{\rotatebox[origin=c]{90}{	$\leftarrow L$, (\ref{eq:joint_loss}) $\rightarrow$}} } & 0.1            & 2.37          & 0.91          & 3.29          & 3.10          & 2.79          & 17.72          & 4.02                & 4.29                & 3.75                & 0.76             \\
\multicolumn{1}{c|}{$L_{SQ}$ only, (\ref{eq:sq_loss})} & 0              & 1.43          & 0.41          & 1.00          & 1.03          & 1.02          & -29.68         & 2.55                & 2.54                & 2.42                & \textbf{0.88}   
\end{tabular}
\end{table*}
The hyperparameter $\alpha$ in (\ref{eq:joint_loss}) is a value between $0$ and $1$ which controls the relative weight of the intrusive and non-intrusive terms which will be analysed in the following.
\section{Experiment 1 - Scaling the Quality Estimator's Influence}\label{sect:exp1}
In this experiment, the \ac{SE} system $\mathcal{G}(\cdot)$ is trained for different $\alpha$ in (\ref{eq:joint_loss}), i.e.~for varying degrees of influence of the quality estimator $\mathcal{D}_B(\cdot)$ in the loss function. In doing this, it is possible to compare the performance of at one pole, a traditional signal-processing-based intrusive loss function, i.e.~$L_\mathrm{spec}$ in (\ref{eq:spec_loss}) only, and at the other a purely non-intrusive \ac{SQ} predictor loss, i.e.~$L_{SQ}$ in (\ref{eq:sq_loss}) only,  as well as points between these poles, i.e.~the combined loss in (\ref{eq:joint_loss}) .  

\subsection{Experiment Setup}
\label{sect:exp1Setup}
Each speech enhancement system model, i.e.~for varying $\alpha$ is trained for $200$ epochs on the VoiceBank-DEMAND~\cite{vb-demand} training set, a widely used dataset for for training speech enhancement networks. It consists of clean English read speech, artificially corrupted with environmental noise from the DEMAND\cite{demand} noise dataset.
The Adam~\cite{kingma2017adam} optimizer is used; following~\cite{dpt-fsnet}, a dynamic strategy to adjust the learning rate is employed, where the learning rate steadily increases during the first few model updates and then scales down over the remaining training epochs.

\subsection{Results}
\autoref{tab:results1} shows the speech enhancement performance of the experiment described in \autoref{sect:exp1Setup} for the VoiceBank-DEMAND testset. The models are evaluated by frequently-used signal-processing-based intrusive measures \ac{PESQ}, \ac{STOI}, the three terms of the Composite measure~\cite{composite} CSIG, CBAK and COVL and the \ac{SI-SDR}~\cite{sisdr}. The models are also evaluated using the non-intrusive neural \ac{SQ} measure \ac{DNSMOS}~\cite{reddy2022dnsmos} as well as in terms of the score assigned by $\mathcal{D}_B(\cdot)$ detailed above in \autoref{sect:predictor}.  

The best performing model in terms of the standard intrusive measures is the model with $\alpha=1$ in \eqref{eq:joint_loss}, i.e.~where no inference of $\mathcal{D}_B(\cdot)$ is used, and the loss function consists solely of the tempo-spectral distance in the loss term defined in \eqref{eq:spec_loss}. Generally, as the value of $\alpha$ decreases, so do the scores. At $\alpha=0$  (i.e.~solely using inference of $\mathcal{D}_B$ as defined in \eqref{eq:sq_loss} as the loss function), the performance degrades significantly, being drastically worse than even the input noisy data in all intrusive measures. The difference in performance between an $\alpha=0$ and $\alpha=0.1$ is stark, suggesting that even a small weighting of the intrusive loss term \eqref{eq:spec_loss} is enough to greatly improve performance.  

Performance assessed by the non-intrusive measures in the right part of \autoref{tab:results1} follows a somewhat different pattern. All models degrade the DNSMOS SIG score in comparison to the noisy (as well as the clean) audio. This is consistent with the findings in masking-based \ac{SE} in general and for the \ac{UDASE} task in particular~\cite{leglaive2024objective}, where all enhancement systems show degraded DNSMOS SIG with the exception of those systems which explicitly optimise towards it in training. While the \ac{DNSMOS} ratings generally decrease as $\alpha$ does, the model for $\alpha = 0.9$ outperforms the model with $\alpha = 1$ in terms of the BAK and OVR components. Furthermore, the model for $\alpha = 0.9$ performs only slightly worse than the model for $\alpha = 1.0$ in terms of the intrusive metrics, suggesting that a small weighting of \eqref{eq:sq_loss} might be beneficial to the overall audio quality. However, this  DNSMOS performance should be interpreted with some scrutiny; the results here show that some of the models outperform even the clean reference audio in terms of DNSMOS, which might be surprising in the first instance. However, later spectrogram analysis (cf.~\autoref{fig:specs}) shows that \emph{clean} signals sometimes contain noise (primarily breathing sounds) which are removed by the \ac{SE} system. Furthermore, all DNSMOS scores for $\alpha = 0$ are much too high given that this system completely destroys the input signal. As noted in \cite{chime7entryclose}, this might be due to an extreme failure to generalise in DNSMOS. 
As it is to be expected, the $\mathcal{D}_B(\cdot)$ score increases as $\alpha$ decreases and inference of $\mathcal{D}_B(\cdot)$ is weighted more heavily in the loss function. 

\subsection{Spectrogram Comparison}
\begin{figure}[!htpb]
\subfloat{%
  \includegraphics[clip,width=.5\columnwidth]{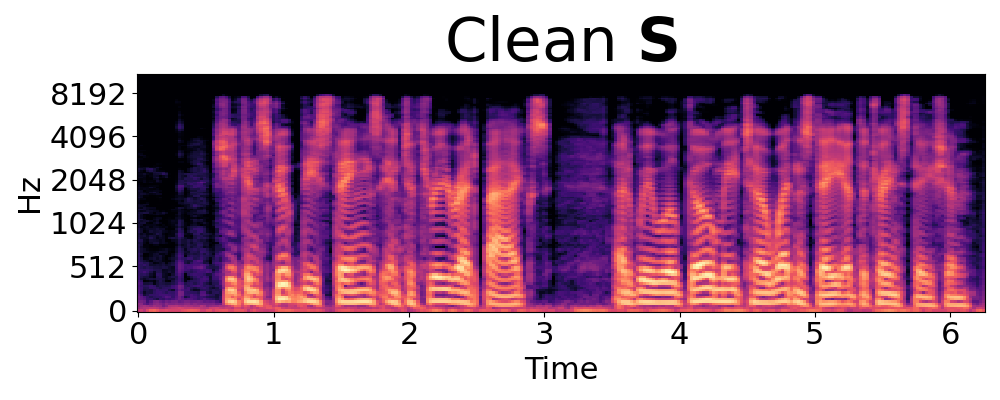}%
}   
\subfloat{%
  \includegraphics[clip,width=.5\columnwidth]{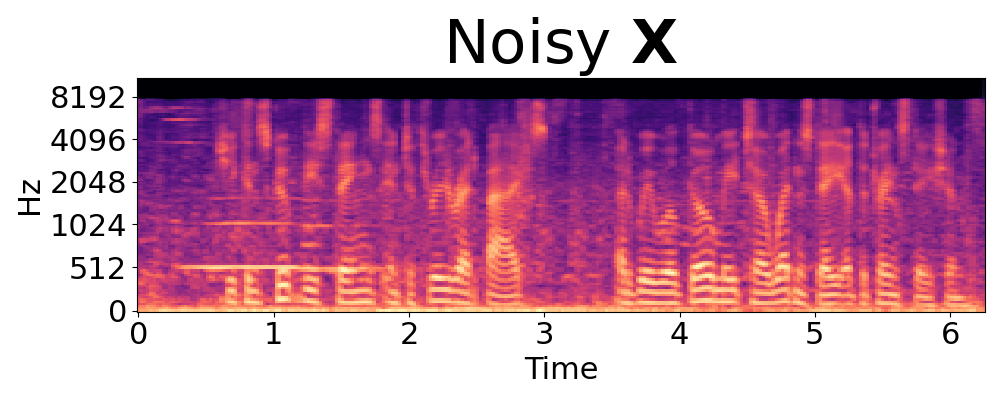}%
}

\subfloat{%
  \includegraphics[clip,width=.5\columnwidth]{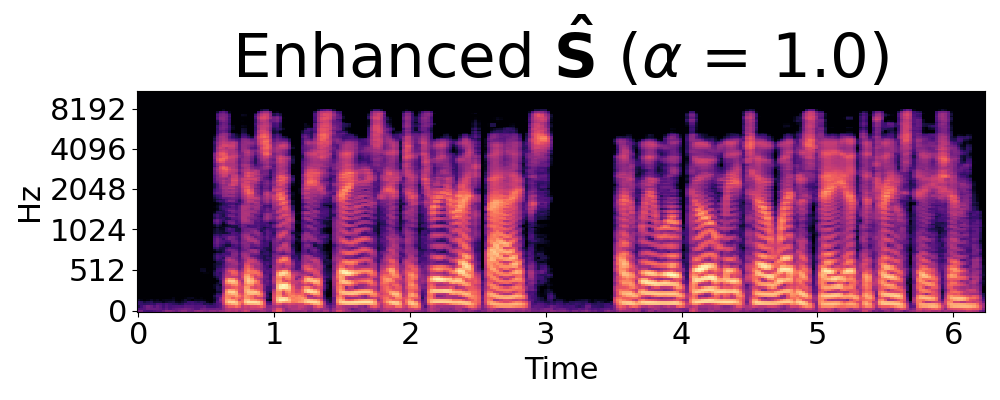}
}
\subfloat{%
  \includegraphics[clip,width=.5\columnwidth]{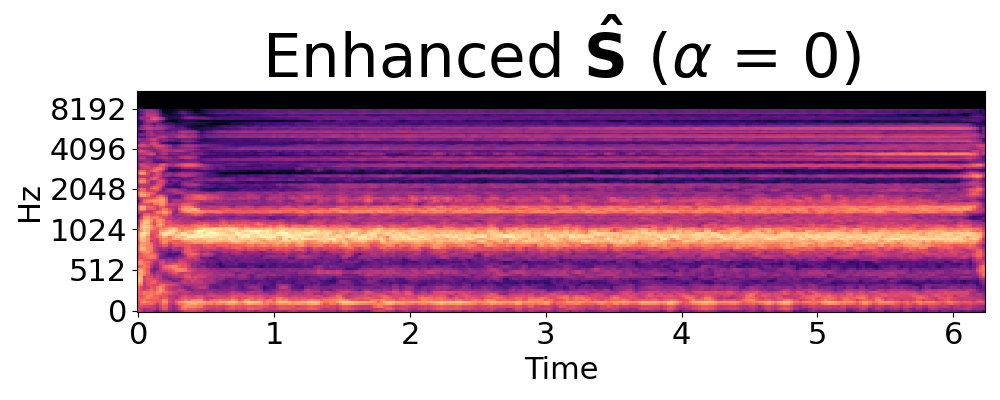}
}

\subfloat{%
  \includegraphics[clip,width=.5\columnwidth]{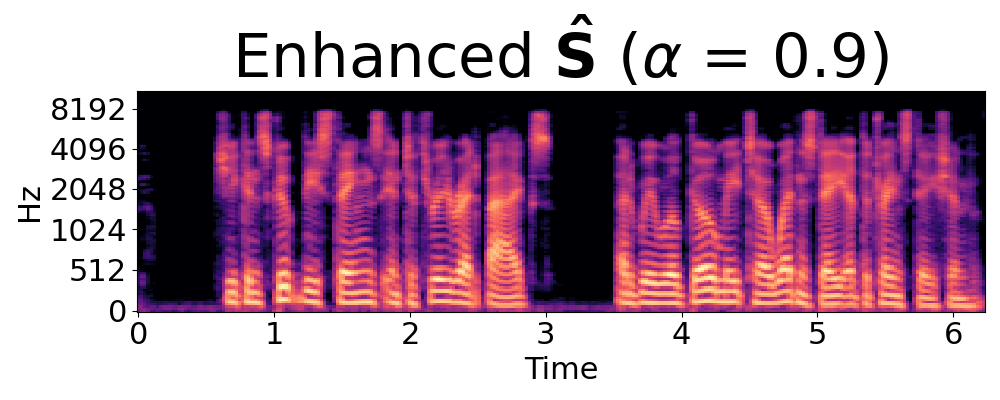}
}
\subfloat{%
  \includegraphics[clip,width=.5\columnwidth]{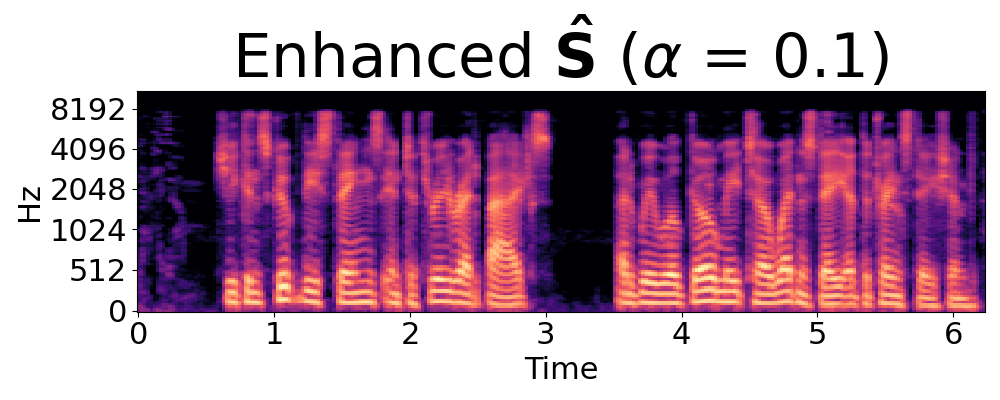}
}

\subfloat{%
  \includegraphics[clip,width=.5\columnwidth]{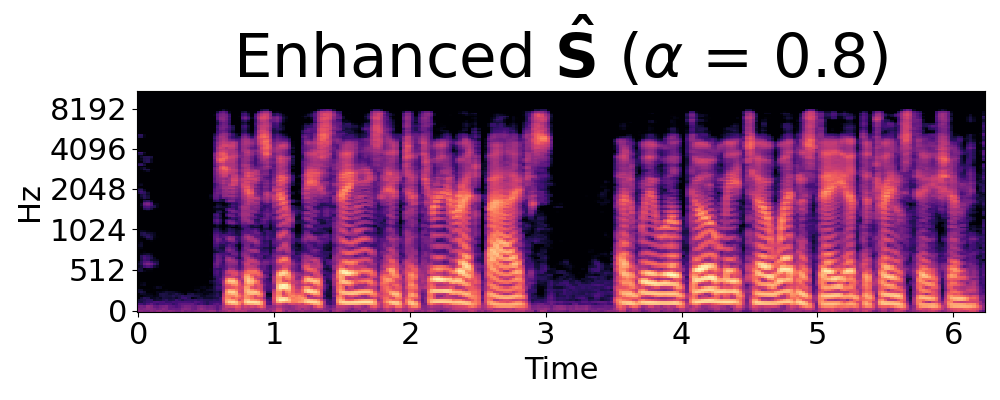}
}
\subfloat{%
  \includegraphics[clip,width=.5\columnwidth]{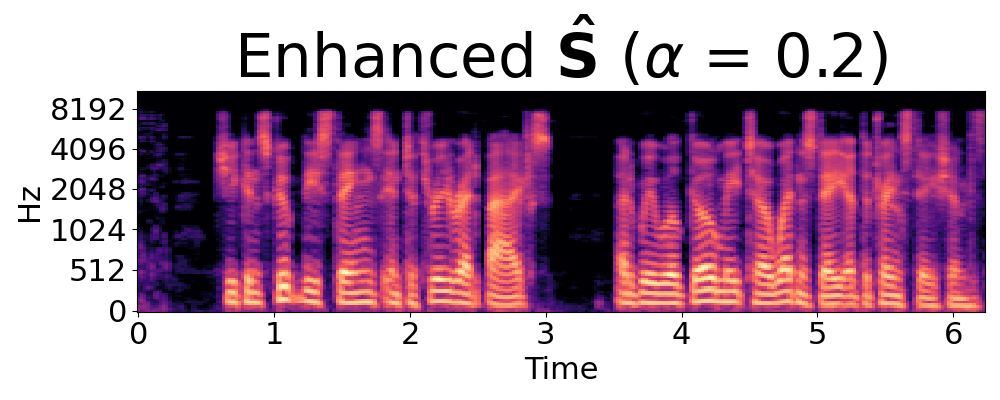}
}

\subfloat{%
  \includegraphics[clip,width=.5\columnwidth]{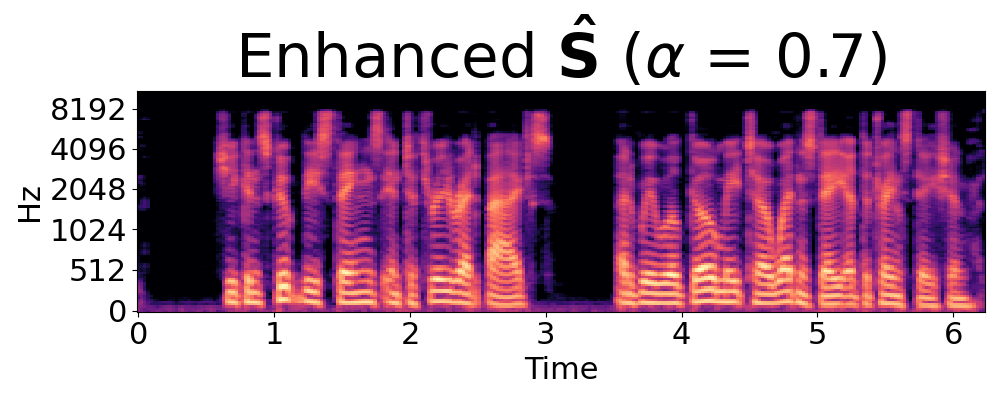}
}
\subfloat{%
  \includegraphics[clip,width=.5\columnwidth]{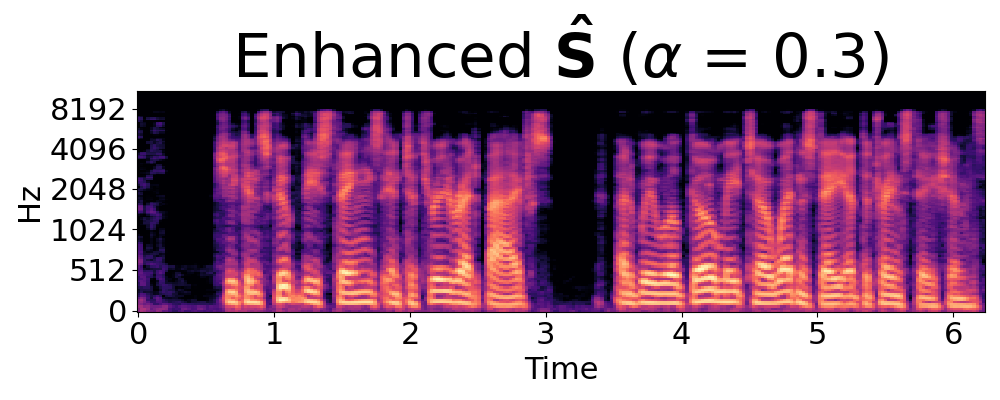}
}

\subfloat{%
  \includegraphics[clip,width=.5\columnwidth]{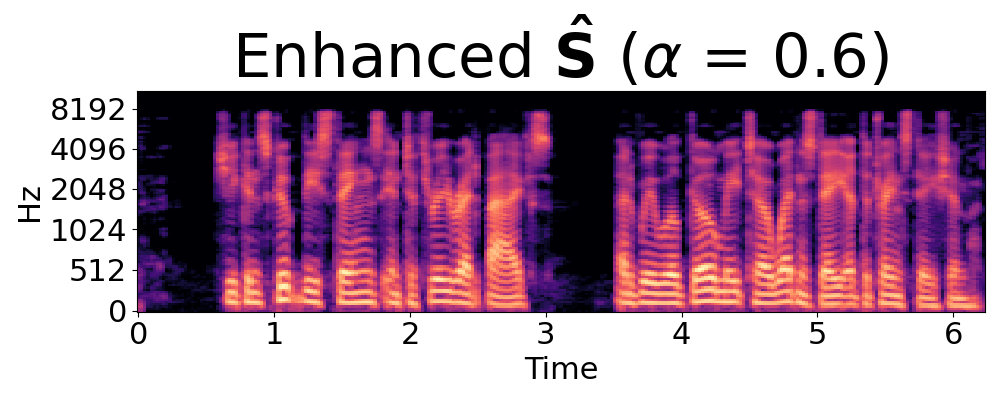}
}
\subfloat{%
  \includegraphics[clip,width=.5\columnwidth]{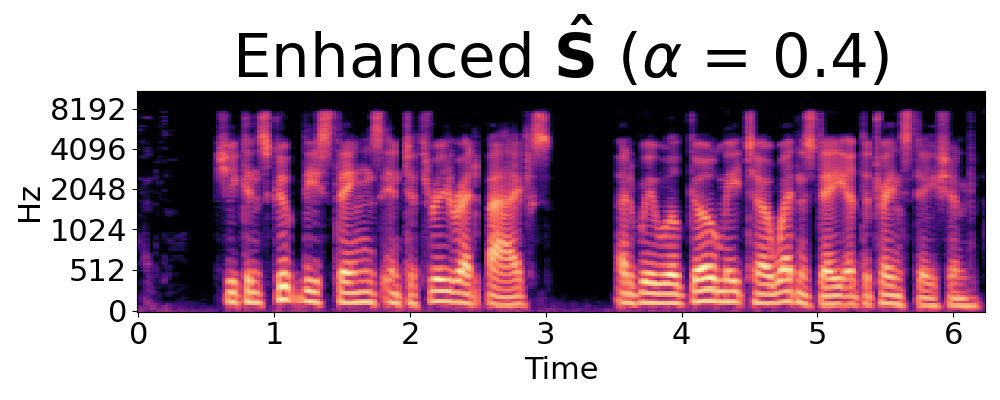}
}

\centering\subfloat{%
  \includegraphics[clip,width=.5\columnwidth]{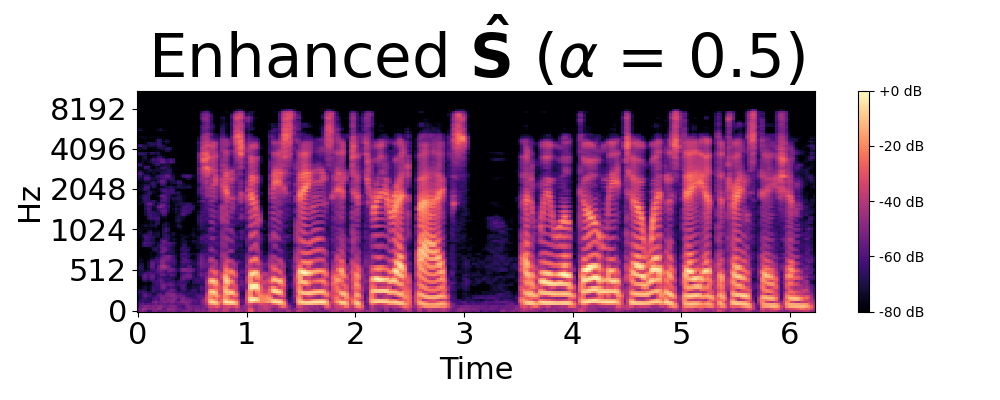}
}
 \caption{(Magnitude) spectrogram comparison for differing values of $\alpha$.}\label{fig:specs}

\end{figure}
\autoref{fig:specs} exemplarily shows spectrograms for the VoiceBank-DEMAND testset file \texttt{p232\_005.wav}; clean reference $\mathbf{S}$ and noisy audio $\mathbf{X}$ (top two panels) as well as enhanced signals $\mathbf{\hat{S}}$ for different $\alpha$ in \eqref{eq:joint_loss} are shown. With the exception of $\alpha = 0$, all models successfully remove the distortion tone present in the first $2$ seconds of the input noisy signal at approx.~$500$~Hz, and generally do enhance the noisy input such that it resembles the clean reference.

As the value of $\alpha$ decreases however, a distortion in the first half second of the audio becomes more prominent. This distortion is interesting for a number of reasons. It does not resemble the noise in this region in the noisy input signal and occurs consistently in appearance spectrally and in audible sound across all audio enhanced by the models, indicating that it can be best characterised as a \textit{hallucination} of the enhancement model(s). This hallucination is most prominent in the model where $\alpha = 0$; other than the hallucination the outputs of this model consist of seemingly meaningless content which does not resemble the noisy input signal at all. Given that the hallucination appears more strongly as the influence of the quality predictor-based loss term \eqref{eq:sq_loss} increases, it is likely caused by the speech enhancement system learning to \textit{trick} $\mathcal{D}_B(\cdot)$. The consistent form of the distortion can also be explained as follows; during the training of $\mathcal{D}_B$, it learned to assign a high-quality rating for input audio which contained a sound like this hallucination. \emph{Then during the training of the \ac{SE} models, 
the \ac{SE} models learn to exploit this quirk of the training of $\mathcal{D}_B$ by introducing the hallucination in order to minimise the loss function.} The consistent temporal position of the hallucination can be explained by the short non-speech region at the start of the audio file which is often present across all audio in the VoiceBank-DEMAND and similar datasets. The presence of this hallucination is likely the cause of the decrease in performance in terms of intrusive signal processing metrics in Table~\ref{tab:results1} while the non-intrusive neural \ac{SQ} metrics change less uniformly; the intrusive metrics all involve a direct comparison with the reference audio which explicitly penalise the presence of the hallucination. The hallucination has a \emph{speech-like} characteristic which is possibly the reason that the \ac{SQ} predictor models reward its presence.  
\section{Experiment 2 - Listening Test }\label{sect:exp2}
In order to better understand the performance of the trained \ac{SE} models and the human perception of the hallucination distortion, a small listening test experiment was carried out.
\subsection{Setup}
Noisy audio files from the VoiceBank-DEMAND testset and audio enhanced by enhancement models with $\alpha$ values of $0$, $0.1$, $0.5$ and $0.9$ were randomly selected for a total of $15$ files ($3$ files from each of the $4$ $\alpha$ values plus the noisy signal). The ITU-T P.835~\cite{ITU_P.835} methodology was used, inspired by \cite{leglaive2023chime7}. $16$ participants sequentially rated each file in terms of the naturalness of the speech signal, the intrusiveness of the background noise distortion and overall quality on $5$ point Likert scales (i.e SIG, BAK and OVRL), for a total of $48$ ratings per audio file. The listening test audio is available online\footnote{\texttt{\url{https://leto19.github.io/nisqa\_se\_demo.html}}}.
\subsection{Results}
The results of the listening test are shown in \autoref{tab:results2}. In terms of signal quality SIG, the noisy input audio scores the highest; this is in line with the \ac{MOS} results reported in \cite{leglaive2023chime7} and results in \autoref{tab:results1}. For BAK and OVRL, the listening-test results follow those of the metrics in \autoref{tab:results1} with the model for $\alpha = 1.0$ being the best performing. 

Interestingly, $\alpha = 0.1$ significantly outperforms $\alpha = 0.5$ in all aspects in the listening tests, outperforming even $\alpha = 0.9$ which showed the best performance in \autoref{tab:results1} in terms of SIG.
The low BAK and OVRL scores of for $\alpha = 0.5$ and $0.1$ suggest that the hallucination is perceptible, but that the listeners considered it as an aspect of the background rather than a distortion in the speech signal itself. This is important when considering the disconnect between intrusive metric scores and human perception \ac{MOS}. An intrusive metric like \ac{PESQ} is \textit{directly comparative} such that deviation in the test signal from the oracle reference signal always results in a lower output score. On the other hand, human \ac{MOS} is \textit{indirectly comparative}; the score is informed wholly by the listener's preconceived notion of speech quality, which varies not only between individuals but also unconsciously over time during the listening test. Likewise, non-intrusive \ac{SQ} predictors are also indirectly comparative, with the output score informed by
the training data.  
This is exemplified clearly by comparing the Composite measure CSIG score of noisy audio in \autoref{tab:results1} with the analogous DNSMOS SIG score in the same table and the real \ac{MOS} SIG average in \autoref{tab:results2}. The noisy signals are generally dissimilar to the clean references, meaning that the intrusive CSIG score suffers but this does not \textit{in reality} mean that the human perception  (or a predictor of human perception) of the speech distortion suffers drastically. 



\begin{table}[!t]
\centering
\caption{Listening Test Results}
Best performance denoted in \textbf{bold}. Unprocessed data denoted in \textit{italic}.
\label{tab:results2}

\begin{tabular}{c|cccccc}
\textbf{}      & \multicolumn{2}{c}{\textbf{SIG}}                                     & \multicolumn{2}{c}{\textbf{BAK}}                                     & \multicolumn{2}{c}{\textbf{OVRL}}                                    \\
$\alpha$ & \multicolumn{1}{l}{\textbf{MEAN}} & \multicolumn{1}{l}{\textbf{STD}} & \multicolumn{1}{l}{\textbf{MEAN}} & \multicolumn{1}{l}{\textbf{STD}} & \multicolumn{1}{l}{\textbf{MEAN}} & \multicolumn{1}{l}{\textbf{STD}} \\ \hline
\textit{noisy}          & \textit{\textbf{4.54}}            & \textit{0.65}                    & \textit{2.92}                     & \textit{0.82}                    & \textit{3.67}                     & \textit{0.88}                    \\
1.0              & 4.50                              & 0.62                    & \textbf{4.67 }                    & 0.66                    & \textbf{4.42}                     & 0.74                             \\
0.9            & 4.31                              & 0.72                             & 4.54                              & 0.65                    & 4.25                              & 0.73                    \\
0.5            & 4.06                              & 0.86                             & 3.58                              & 1.03                             & 3.63                              & 0.96                             \\
0.1            & 4.35                              & 0.70                    & 3.83                              & 0.69                             & 3.94                              & 0.81                           
\end{tabular}
\end{table}

\section{Conclusion}
\label{sec:Conclusion}
\ac{SE} models which are optimised using non-intrusive neural \ac{SQ} predictors are shown to produce hallucinatory artefacts in output audio. These hallucinations do not represent meaningful content but are learned by the \ac{SE} system in order to optimise the audio towards maximising the score awarded by the \ac{SQ} predictor. Intrusive metrics like \ac{PESQ} are sensitive to these hallucinations, and they are shown to generally be perceptible in human listening tests. 
\bibliographystyle{IEEEtran}
\bibliography{refs}
\end{document}